\documentclass[preprints,article,accept,pdftex,moreauthors]{Definitions/mdpi} 
\usepackage{amsmath}
\usepackage{amssymb}
\usepackage{graphicx}
\usepackage{color}
\usepackage{subfig}
\usepackage{bm}
\usepackage{gensymb}
\usepackage{bigints}
\usepackage{pifont}
\usepackage{aas_macros}
\usepackage{physics}
\firstpage{1} 
\pubvolume{1}
\issuenum{1}
\articlenumber{0}
\pubyear{2022}
\copyrightyear{2022}
\datereceived{} 
\dateaccepted{} 
\datepublished{} 
\hreflink{https://doi.org/} 
\pdfoutput=1

\Title{Cooling Process of White Dwarf Stars in Palatini $f(R)$ Gravity}

\TitleCitation{Cooling Process of White Dwarf Stars in Palatini $f(R)$ Gravity}

\Author{Surajit Kalita$^{1}$\orcidA{}, Lupamudra Sarmah$^{2,3}$\orcidB{} and Aneta Wojnar$^{4}$\orcidC{}}

\AuthorNames{Surajit Kalita, Lupamudra Sarmah and Aneta Wojnar}

\AuthorCitation{Kalita, S., Sarmah, L. and Wojnar, A.}

\address{%
$^{1}$ \quad High Energy Physics, Cosmology \& Astrophysics Theory (HEPCAT) Group, Department of Mathematics \& Applied Mathematics, University of Cape Town, Cape Town 7700, South Africa; surajit.kalita@uct.ac.za\\
$^{2}$ \quad Indian Institute of Astrophysics, Bengaluru 560034, India; lupamudra.sarmah@iiap.res.in\\
$^{3}$ \quad Department of Physics, Indian Institute of Technology (BHU), Varanasi 221005, India\\
$^{4}$ \quad Laboratory of Theoretical Physics, Institute of Physics, University of Tartu, W. Ostwaldi 1, Tartu 50411, Estonia; aneta.magdalena.wojnar@ut.ee}

\corres{Correspondence: surajit.kalita@uct.ac.za}

\abstract{A simple cooling model of white dwarf stars is re-analyzed in Palatini $f(R)$ gravity. Modified gravity affects the white dwarf structures and consequently their ages. We find that the resulting super-Chandrasekhar white dwarfs need more time to cool down than sub-Chandrasekhar ones, or when compared to the Newtonian models.}


\begin{document}


\section{Introduction}

Gravity, one of the most fundamental interactions that are readily experienced in everyday life, remains enigmatic. To date, general relativity~(GR) continues to be a successful and well-accepted theory for explaining gravitational phenomena. Fulfilling the shortcomings of Newtonian gravity, GR can quite accurately explain a plethora of phenomena, starting from the perihelion precession of Mercury and gravitational redshift to the prediction of gravitational waves~\cite{2004sgig.book.....C}. Even though Einstein's theory of gravity is well-tested in the weak-field regime with remarkable accuracy~\cite{poisson2014gravity}, it might break down in the very strong gravity regime. This, along with recent cosmological evidences~\citep{1999ApJ...517..565P,1998AJ....116.1009R}, suggests that GR may not be adequate to explain the universe and demands appropriate modifications to this theory. Several cosmological observations indicate that ordinary matter comprises only a tiny percentage of the total energy component of the universe and is inadequate to explain the two phases of cosmic acceleration~\citep{1981PhRvD..23..347G,2003A&G....44a..32C,1998AJ....116.1009R,1999ApJ...517..565P,1999PhRvD..60h1301H}. In order to make GR compatible with these cosmological data and other observations related to the rotation curve of the galaxies~\cite{2000FCPh...21....1B}, dark matter and dark energy were postulated to exist~\cite{2016ConPh..57..496G,2008JCAP...03..022T}. The popular cosmological model, viz. $\Lambda$CDM model~\cite{bullock2017small}, can fit the new observations with remarkable accuracy. However, this model is burdened with well-known cosmological problems and is unable to explain the origin of inflation or dark matter \cite{2017Galax...5...17D}. Although the nature and origin of dark matter still remain a mystery, a scalar field with a slowly varying potential can be considered as a candidate for inflation as well as dark energy~\cite{1999AN....320...97G,2022PhRvD.105h3533G}. On the other hand, a pure geometric contribution also seems to able to play the same role as it does the scalar field~\cite{borowiec2012cosmic,koivisto2006cosmological,flanagan2004palatini,fay2007f,sotiriou2006unification,Dioguardi:2021fmr,Gialamas:2020snr,Dimopoulos:2022rdp,Szydlowski:2015fcq,Szydlowski:2017uuy,Stachowski:2016zio,Allemandi:2005qs,Allemandi:2004wn,Allemandi:2004ca}. 

The last approach means that one modifies Einstein's theory of gravity~\cite{2011PhR...509..167C,2021mgca.book.....S,2018JCAP...10..054H,2007PhRvD..75h4031F} in order to solve the aforementioned problems. Though there are several ways of extending and modifying GR, the simplest one is the $f(R)$ gravity \cite{1970MNRAS.150....1B,2010RvMP...82..451S}, in which the Lagrangian density is considered to be an arbitrary function of the scalar curvature $R$. In order to obtain the modified field equations, one can apply either of the two variational principles to the action: the standard metric variation or the Palatini variation. The former approach involves varying the action with respect to the metric only, leading to the fourth-order field equations for the metric\footnote{Although, one can rewrite the equations as second-order differential equations for the metric, and the additional one for the curvature scalar, which arises to a dynamical field in this theory.}. On the other hand, the latter is the variation of the action with respect to the metric as well as the connection, waiving thus the assumption that those two objects are related to each other, as it happens in GR or in the metric formulation. It gives rise to the second-order field equations~\citep{2007PhRvD..75f3509F,2006CQGra..23.1253S}. It is worth mentioning that both approaches are equivalent only in the context of GR when $f(R)$ is a linear function of $R$. Otherwise, for a general $f(R)$ term in the action, they might lead to entirely different theories and spacetime structures. This is so because the connection in the Palatini approach depends on the particular $f(R)$ form\footnote{This fact arises as a conclusion from the field equations and it will be evident in the next section.}, whereas in the metric formalism, the connection is assumed to be the Levi-Civita one of the spacetime metric. 
Moreover, being more general in nature and comparatively easier to work with, the Palatini $f(R)$ gravity is, therefore, being increasingly used to study a wide variety of phenomena in recent times. Apart from the cosmological models based on the Palatini $f(R)$ gravity to study cosmic acceleration~\citep{2006PhRvD..73f3515S,2004GReGr..36.1765N,2006A&A...454..707A,2016JCAP...01..040B,2020PhRvD.102d4029J}, several astrophysical systems including white dwarfs (WDs), neutron stars \cite{2017GReGr..49...25T,2021EPJC...81..888H}, stellar \cite{Olmo:2019qsj,2020PhRvD.102l4045W,2021PhRvD.103d4037W} and substellar objects~\citep{2021PhRvD.103f4032B,2021PhRvD.104j4058W,Wojnar:2022ttc,Kozak:2021ghd,Kozak:2021zva,Kozak:2021fjy} have been extensively studied using this theory (for review, see~\cite{Olmo:2019flu,Wojnar:2022txk}).

WDs represent the final evolutionary stage of progenitor stars with masses less than $(10\pm 2) M_\odot$~\cite{1986bhwd.book.....S,2018MNRAS.480.1547L}. Although the less massive WDs with mass being less than about $0.4 M_\odot$, seem to be predominantly composed of helium, the other WDs, in general, might contain heavier elements like carbon and oxygen~\citep{2018MNRAS.480.1547L}. Moreover, in all such stars, the core is surrounded by a thin layer of helium, which in turn is surrounded by an even thinner layer of hydrogen. In order to support this structure of WDs, the inward gravitational force is balanced by the electron degeneracy pressure acting outwards due to the Pauli exclusion principle. Since these stars cannot derive energy from thermonuclear reactions, their evolution can be described simply as a cooling process. The degenerate core of the WDs acts as a reservoir of energy, whereas the outer non-degenerate layers are responsible for energy outflows. The simplest models assume that WDs possess temperatures substantially lower than the Fermi temperature. As a result, zero temperature calculations seem to suffice to obtain their structures, however temperature does play an important role to obtain their ages. By studying cooling processes, one can calculate the lifetime of these stars, which can provide important information about the age of different galactic components, stellar formation rate, and past galactic history. 

In recent times, with the observations of various over- and under-luminous type Ia supernovae, WDs are extensively being studied in the framework of $f(R)$ gravity~\citep{2018JCAP...09..007K,2021ApJ...909...65K,2021IJGMM..1840006W,2022PhLB..82736942K,2022PhRvD.105b4028S,2015JCAP...05..045D} and other theories of gravity or fundamental physics~\citep{das2015imprint,garcia1995rate,althaus2011evolution,corsico2013independent,benvenuto2004asteroseismological,2018JCAP...05..028S,2019JCAP...02..040L,2017EPJC...77..871C,2019PhRvD..99j4074E,biesiada2004new,benvenuto1999evolution,bienayme2002white,2002EAS.....2..123I,2016CQGra..33w5014B,2019PhRvD.100b4025C,wibisono2018information,2002PhRvD..65d3008B,panah2019white,panotopoulos2020white,isern2003white,2003NuPhS.114..107I,2017JCAP...10..004B}. Other than a violation of the Chandrasekhar limit, we also expect a modification in the physical properties of the WDs, including the cooling rate, and hence their ages. Therefore, this work aims at studying a part of the cooling process in WDs using Palatini $f(R)$ gravity. This will allow to estimate the modifications in their lifetime, brought in by the particular theory of gravity.

The following is a breakdown of how this article is structured. In Section~\ref{Sec2}, we discuss the basic formalism of Palatini $f(R)$ gravity and briefly review how it modifies the hydrostatic balance equations. We further employ these equations in Section~\ref{Sec3} to derive the modified temperature gradient equation in Palatini $f(R)$ gravity and thereby obtain the modified cooling age of WDs. In Section~\ref{Sec4}, we numerically calculate the masses and radii of the modified gravity inspired WDs and their ages. Finally, in Section~\ref{Sec5}, we put forward our concluding remarks on this work.

\section{Basic formalism of Palatini $f(R)$ gravity and hydrostatic balance equations}\label{Sec2}

In this paper, the metric signature convention is $(-,+,+,+)$. For a spacetime metric $g_{\mu\nu}$, the action for $f(R)$ gravity has the following form~\cite{2010LRR....13....3D}
\begin{align}
    S[g,\Gamma,\Psi]=\frac{1}{2\mathtt{k}}\int\sqrt{-g}f\big(R(g,\Gamma)\big)\dd[4]{x} + S_\mathrm{m}[g,\Psi],
\end{align}
where $\mathtt{k}=-8\pi G/c^4$, $g=\det(g_{\mu\nu})$, $G$ is Newton's gravitational constant, $c$ is the speed of light, and $S_\mathrm{m}$ is the matter action depending on the spacetime metric and matter fields $\Psi$ only. On the other hand, the Ricci-Palatini curvature scalar is constructed with two independent objects: the connection $\Gamma$ and the metric $g_{\mu\nu}$, such that $R=R_{\mu\nu}(\Gamma)g^{\mu\nu}$. 

Varying this action with respect to $g_{\mu\nu}$ results in the modified field equation, given by~\citep{2010LRR....13....3D}
\begin{equation}\label{Eq: f(R) master eq}
    f'(R)R_{\mu\nu}-\frac{1}{2}f(R)g_{\mu\nu}=\mathtt{k} \mathcal{T}_{\mu\nu},
\end{equation}
where $f'(R) = \dv*{f(R)}{R}$ and $\mathcal{T}_{\mu\nu}$ is the energy-momentum tensor defined as usually  $\mathcal{T}_{\mu\nu}=-\left(2/\sqrt{-g}\right)\delta S_m/\delta g^{\mu\nu}$. Let us notice that for the spherical-symmetric WD in the statistical equilibrium, the microscopic description of Fermion gas comes down to the perfect-fluid description with the Chandrasekhar equation of state \cite{chavanis2020statistical,2022arXiv220804023W}. Moreover, the Ricci-Palatini tensor $R_{\mu\nu}$ must be symmetric, since its antisymmetric part introduces instabilities \cite{Borowiec:1996kg,Allemandi:2004wn,BeltranJimenez:2019acz}.

In contrast, varying the action with respect to $\Gamma$ and performing some small algebraic transformation, gives the following equation:
\begin{equation}
    \nabla_{\lambda}\left(\sqrt{-g}f'(R)g^{\mu\nu}\right)=0,
\end{equation}
where $\nabla_{\lambda}$ is the covariant derivative ruled by $\Gamma$. Now, defining a new metric tensor $\bar g_{\mu\nu} = f'(R)g_{\mu\nu}$, the above equation can be recast as
\begin{equation}
    \nabla_\lambda(\sqrt{-\bar g}\bar g^{\mu\nu})=0,
\end{equation}
providing that $\Gamma$ is the Levi-Civita connection with respect to $\bar g_{\mu\nu}$. Let us notice that it is conformally related to the spacetime metric $g$, allowing us to rewrite the field equations in a much simpler form in order to perform rather tedious calculations \cite{afonso2018mapping,afonso2018mapping1,afonso2019correspondence}. Moreover, the trace of Equation~\eqref{Eq: f(R) master eq} with respect to $g_{\mu\nu}$ is given by
\begin{equation}\label{Eq: trace equation}
    f'(R)R-2f(R)=\mathtt{k}\mathcal{T},
\end{equation}
where $\mathcal{T}=g^{\mu\nu}\mathcal{T}_{\mu\nu}$. It is clear now that the curvature is not dynamical, and it can be expressed by the trace of the energy-momentum tensor for a given functional $f(R)$. In other words, all the modifications introduced by Palatini $f(R)$ gravity are functions of the matter fields.

In this paper, we work with the Starobinsky model of $f(R)$ gravity, i.e., $f(R)=R+\alpha R^2$ with $\alpha$ being the model parameter. Hence, the trace equation reduces to the relation between the scalar curvature and the trace of the energy momentum tensor, which resembles the one known from GR, given by
\begin{equation}
    R=-\mathtt{k} \mathcal{T}.
\end{equation}

In our previous study \cite{2022PhRvD.105b4028S}, we analyzed hydrostatic equilibrium equations in both Jordan and Einstein frames separately. We use the Einstein frame's equations to perform rather tedious calculations in the case of the Jordan ones, and then we transform them back to the Jordan frame, in which we are equipped with the equation of state. The frames are related by the conformal transformation $\bar g_{\mu\nu} = f'(R)g_{\mu\nu}$. Thereby we obtained the structure of WDs and later analyzed their stabilities with respect to the radial perturbations. We considered the weak-gravity regime for those equations\footnote{To see the relativistic hydrostatic equilibrium equation for Palatini $f(R)$ gravity, see \cite{Wojnar:2017tmy}.} in a spherically symmetric case, that is, the pressure $\Tilde P$ is insignificant with respect to the density $\tilde\rho$: $\tilde P(\tilde r)\ll\tilde\rho(\tilde r) c^2$ with $4\pi \Tilde{r}^3\tilde P/c^2\ll m(\tilde r)$, as well as for the geometric contribution, that is, $2Gm(\tilde r)/\Tilde{r}c^2\ll 1$, where $m$ is the mass function. The `tilde' denotes variables in the Einstein frame which are related to the physical ones by the tranformations $\Tilde{r}^2 = \Phi r^2$, $\Tilde{\rho}= \rho/\Phi^2$, and $\Tilde{P}= P/\Phi^2$, where $\Phi := f'(R)$. Hence, for the considered Starobinsky model, we have
\begin{equation}
     \Phi  = 1+2\alpha R = 1 + 2\alpha \mathtt{k} \rho c^2.
\end{equation}
Thus, when we take into account the above simplifications, the hydrostatic balance equations in the Einstein frame, are given by 
\begin{align}\label{Eq: dPdr Einstein}
    \dv{\tilde P}{\Tilde{r}} &= -\frac{G m \tilde\rho}{\Phi \Tilde{r}^2},\\ 
    \label{Eq: dmdr Einstein}
    \dv{m}{\Tilde{r}} &= 4\pi \Tilde{r}^2 \tilde\rho.
\end{align}
Now, performing the confromal transformation, one can obtain the hydrostatic balance equations in the Jordan frame, given by
\begin{align}\label{Eq: dPdr Jordan}
    \dv{P}{r} &= -\frac{Gm\rho}{\Phi^{\frac{3}{2}} r^2}\left(1+\frac{1}{2}r\frac{\Phi'}{\Phi}\right) + 2P\frac{\Phi'}{\Phi},\\
    \label{Eq: dmdr Jordan}
    \dv{m}{r} &= \frac{4\pi r^2\rho}{\Phi^{\frac{1}{2}}} \left(1+\frac{1}{2}r\frac{\Phi'}{\Phi}\right),
\end{align}
where we define $\Phi' = \dv*{\Phi}{r} = 2\alpha\mathtt{k}c^2\dv*{\rho}{r}$. 

In order to solve the above system of differential equations, one needs to choose an equation of state (EoS). It relates the microscopic variables defining a physical system, as well as carries information about additional forces between particles, dependence on the temperature or phase transitions points, to mention just some of them. We consider a quite simple EoS, that is, the barotropic EoS of the form $P=P(\rho)$, neglecting its dependence on the temperature and other thermodynamic variables. This is so because, in a WD, temperature is much lower than the Fermi temperature due to its high density. Since in this paper, we focus on the non-rotating and non-magnetized WD stars, which can be well modelled as spherical-symmetric balls consisting of degenerate electrons, the EoS describing the microscopic properties of such a system is given by the Chandrasekhar one, written in the parametric form as~\cite{1935MNRAS..95..207C}
\begin{align}
    P &= \frac{\pi m_\text{e}^4 c^5}{3 h^3}\left[x_\text{F}\left(2x_\text{F}^2-3\right)\sqrt{x_\text{F}^2+1}+3\sinh^{-1}x_\text{F}\right], \nonumber\\ \label{Chandrasekhar EoS}
    \rho &= \frac{8\pi \mu_\text{e} m_\text{H}(m_\text{e}c)^3}{3h^3}x_\text{F}^3,
\end{align}
where $x_\text{F} = p_\text{F}/m_\text{e}c$ with $p_\text{F}$ being the Fermi momentum, $m_\text{e}$ the mass of an electron, $h$ the Planck's constant, $\mu_\text{e}$ the mean molecular weight per electron, and $m_\text{H}$ the mass of a hydrogen atom. Solving Equations~\eqref{Eq: dPdr Jordan} and~\eqref{Eq: dmdr Jordan} simultaneously along with this EoS, one can obtain the mass--radius relation of WDs in the considered model of gravity.

\section{Temperature gradient equation and cooling timescale of white dwarfs in $f(R)$ gravity}\label{Sec3}

Let us consider a very simple model of cooling process. In what follows, we assume that the WD star radiates its energy away and does not have any other energy sources. Therefore, its cooling process depends only on the atmosphere properties with the energy transports through the star and the model of gravity. In case of GR, a similar analysis is given in~\cite{1986bhwd.book.....S,1952MNRAS.112..583M}.

Moreover, we assume the WD behaves as a perfect black body, such that the equation for radiative energy transport in Einstein frame is given by
\begin{align}\label{Eq: dTdr Einstein}
    \dv{T}{\Tilde{r}} = -\frac{3L\kappa\tilde\rho}{4\pi\Tilde{r}^2 4acT^3},
\end{align}
where $T(\Tilde{r})$ is the temperature of the WD at a radius $\Tilde{r}$, $L$ is the luminosity at that radius, $\kappa$ is the opacity, and $a = 8\pi^5 k_\mathrm{B}^4/15c^3h^3 \approx 7.6\times10^{-15}\rm\,erg\,cm^{-3}\,K^{-4}$ is the radiation constant with $k_\mathrm{B}$ being the Boltzmann constant. Moreover, for the perfect black body, spatial variation of $L$ is given by
\begin{equation}
    \dv{L}{\Tilde{r}} = 4\pi\Tilde{r}^2\tilde\rho(\Tilde{r})\epsilon(\Tilde{r}),
\end{equation}
where $\epsilon$ is the power produced per unit mass of stellar material. Now, combining Equations~\eqref{Eq: dPdr Einstein} and~\eqref{Eq: dTdr Einstein}, we obtain
\begin{align}
    \pdv{T}{\tilde P} = \frac{3L\kappa\Phi}{16\pi ac G m T^3}.
\end{align}
Thus, defining $\nabla = \pdv*{\ln{T}}{\ln{\tilde P}}$, we obtain
\begin{align}
    \tilde\nabla = \frac{3L\kappa\tilde P\Phi}{16\pi ac G m T^4}.
\end{align}

In Jordan frame, using Equation~\eqref{Eq: dTdr Einstein}, the temperature gradient equation is given by
\begin{align}\label{Eq: dTdr Jordan}
    \dv{T}{r} = -\frac{3L\kappa\rho}{4\pi r^2 4acT^3 \Phi^{5/2}}\left(1+\frac{1}{2}r\frac{\Phi'}{\Phi}\right).
\end{align}
Therefore, combining Equations~\eqref{Eq: dPdr Jordan} and~\eqref{Eq: dTdr Jordan}, we obtain
\begin{align}\label{Eq: dTdP}
    \pdv{T}{P} = \frac{3L\kappa}{16\pi ac G m T^3\Phi} - \frac{3L\kappa\rho\left(1+\frac{1}{2}r\frac{\Phi'}{\Phi}\right)}{32\pi ac r^2 T^3 P\Phi^{3/2}\Phi'},
\end{align}
and thereby\footnote{Let us notice that this form differs slightly from the one obtained in~\cite{2020PhRvD.102l4045W}. This is so because of different assumptions on the matter description and its behaviour under the conformal transformation.}
\begin{align}
    \nabla = \pdv{\ln{T}}{\ln{P}} = \frac{3L\kappa P}{16\pi ac G m T^4\Phi} - \frac{3L\kappa\rho\left(1+\frac{1}{2}r\frac{\Phi'}{\Phi}\right)}{32\pi ac r^2 T^4 \Phi^{3/2}\Phi'}.
\end{align}
This factor $\nabla$ determines the dynamical stability against convective processes in the modified gravity inspired stars~\cite{2020PhRvD.102l4045W}. For a fluid parcel inside a star, if the $\nabla<\nabla_\text{ad}$, where $\nabla_\text{ad}$ the adiabatic gradient, it is convectively stable. This is famously known as the Schwarzschild stability condition. For a degenerate gas following $P\propto\rho^\Gamma$, $\nabla_\text{ad} = 1-1/\Gamma$. Near the surface of a WD, generally the EoS is non-relativistic with $\Gamma=5/3$ (can be obtained from Equation~\eqref{Chandrasekhar EoS} when the non-relativistic limit is taken), and hence in this case, $\nabla_\text{ad}$ turns out to be 2/5.

We assume Kramer's opacity throughout the paper, i.e., $\kappa = \kappa_0\rho T^{-3.5}$ with $\kappa_0 = 4.34\times 10^{24} Z \left(1+X\right)\rm\,cm^2\,g^{-1}$, where $X$ is the mass-fraction of hydrogen and $Z$ is the mass-fraction of metals (elements other than hydrogen and helium). Therefore, from Equation~\eqref{Eq: dTdP}, we have
\begin{align}\label{Eq: dPdT}
    \dv{P}{T} = \frac{16\pi ac G m T^{6.5}\Phi}{3L\kappa_0\rho} - \frac{32\pi ac r^2 T^{6.5} P\Phi^{3/2}\Phi'} {3L\kappa_0\rho^2 \left(1+\frac{1}{2}r\frac{\Phi'}{\Phi}\right)}.
\end{align}
Near the surface, it is reasonable to assume the ideal gas EoS
\begin{align}\label{Eq: ideal gas}
    P = \frac{\rho k_\mathrm{B} T}{\mu m_\mathrm{u}},
\end{align}
where $m_\mathrm{u}$ is the atomic mass unit and $\mu$ is the mean molecular weight. Moreover, near the surface, $m$ can be replaced by the total mass of the WD, that is, $m(r\approx\mathcal{R})=M$. Therefore, combining Equations~\eqref{Eq: dPdT} and~\eqref{Eq: ideal gas}, we obtain
\begin{align}
    P\dv{P}{T} = \frac{16\pi ac G M k_\mathrm{B} T^{7.5}\Phi}{3L\kappa_0\mu m_\mathrm{u}} - \frac{32\pi ac \mathcal{R}^2 k_\mathrm{B}^2T^{8.5} \Phi^{3/2}\Phi'} {3L\kappa_0\mu^2 m_\mathrm{u}^2 \left(1+\frac{1}{2}\mathcal{R}\frac{\Phi'}{\Phi}\right)}.
\end{align}
Integrating this equation with the boundary condition $P=0$ at $T=0$, we obtain
\begin{align}
    P &= \left(\frac{2}{8.5} \frac{4ac\Phi}{3} \frac{4\pi G M}{\kappa_0 L}\right)^{1/2}\left(\frac{k_\mathrm{B}}{\mu m_\mathrm{u}}\right)^{1/2} T^{4.25} - \left(\frac{4}{9.5} \frac{4ac\Phi^{3/2}\Phi'}{3\left(1+\frac{1}{2}\mathcal{R}\frac{\Phi'}{\Phi}\right)}\frac{4\pi \mathcal{R}^2} {\kappa_0 L}\right)^{1/2} \left(\frac{k_\mathrm{B}}{\mu m_\mathrm{u}}\right) T^{4.75}.
\end{align}
Equating this pressure with the ideal gas pressure in Equation~\eqref{Eq: ideal gas}, we obtain
\begin{align}\label{Eq: rho_T relation}
    \rho &= \left(\frac{2}{8.5} \frac{4ac\Phi}{3} \frac{4\pi G M}{\kappa_0 L} \frac{\mu m_\mathrm{u}}{k_\mathrm{B}}\right)^{1/2} T^{3.25} - \left(\frac{4}{9.5} \frac{4ac\Phi^{3/2}\Phi'}{3\left(1+\frac{1}{2}\mathcal{R}\frac{\Phi'}{\Phi}\right)}\frac{4\pi \mathcal{R}^2} {\kappa_0 L}\right)^{1/2} T^{3.75}.
\end{align}
An order of magnitude calculation for a WD shows that the first term on the right hand side is at least $3-4$ orders of magnitude larger than the second one. Thus, for further calculations, we can safely drop the second term. Near the surface, the non-relativistic EoS is followed, and hence in Equation~\eqref{Chandrasekhar EoS}, assuming $x_\text{F}\ll1$, we obtain $P \approx 1.0\times10^{13} \left(\rho/\mu_\text{e}\right)^{5/3}$. Substituting $P$ in the ideal gas EoS~\eqref{Eq: ideal gas}, we obtain the density $\rho_*$ and temperature $T_*$ at the interface radius inside which temperature nearly remains constant. Thus equating degenerate pressure and ideal gas pressure, we obtain $\rho_* = 2.4\times10^{-8} \mu_\mathrm{e} T_*^{3/2}\rm\,g\,cm^{-3}$. Substituting it in Equation~\eqref{Eq: rho_T relation} and after some rearrangements, luminosity is given by
\begin{align}\label{lum1}
    L \approx \left(5.8\times 10^5 \rm\, erg\,s^{-1}\right) \frac{\mu}{\mu_\mathrm{e}^2}\frac{\Phi}{Z(1+X)}\frac{M}{M_\odot} T_*^{3.5}.
\end{align}
Assuming a carbon-oxygen WD with the surface containing $90\%$ of helium and $10\%$ other heavy elements, we have $\mu_\text{e}=2$, $X=0$, $Y (\text{helium mass-fraction})=0.9$, $Z=0.1$, and $\mu=1.4$. Thus the above luminosity expression reduces to
\begin{align}\label{Eq: Luminosity}
    L = \left(2\times 10^6 \rm\, erg\,s^{-1}\right) \Phi\frac{M}{M_\odot}T_*^{3.5}.
\end{align}

Because luminosity is defined as the energy $E$ radiated per unit time $t$, it is given by
\begin{equation}
    L = -\dv{E}{t}.
\end{equation}
Defining a constant $\mathcal{C}$ such that $\mathcal{C} M_\odot = 2\times 10^6 \rm\, erg\,s^{-1}$ and substituting $L$ from Equation~\eqref{Eq: Luminosity}, we obtain
\begin{align}
    -\dv{t}(\frac{3k_\mathrm{B}TM}{2Am_\mathrm{u}}) =  \mathcal{C}M\Phi T_*^{7/2}.
\end{align}
Integrating this equation from time $t_0$ to the present time $t$ such that $T_0$ was the initial temperature and $T_*$ is present temperature of the WD, we obtain
\begin{align}
    \frac{3k_\mathrm{B}}{5Am_\mathrm{u}}\left(T_*^{-5/2}-T_0^{-5/2}\right) = \mathcal{C}\Phi\left(t-t_0\right).
\end{align}
Assuming $T_0\gg T_*$ and defining cooling time as $\tau = t-t_0$ and denoting $L_*$ is the present surface luminosity of the WD, the use of the Equation~\eqref{Eq: Luminosity} allows us to approximate the cooling timescale as a function of mass and luminosity for a WD as follows:
\begin{align}
    \tau &= \frac{3}{5}\frac{k_\mathrm{B}T_* M}{Am_\mathrm{u}L_*} 
    = \frac{3}{5}\frac{k_\mathrm{B}}{Am_\mathrm{u}\mathcal{C}\Phi T_*^{5/2}}
    = \frac{3}{5}\frac{k_\mathrm{B}}{Am_\mathrm{u} \mathcal{C}^{2/7} \Phi^{2/7}}\left(\frac{M}{L_*}\right)^{5/7}.\label{Eq: age3}
\end{align}
In the next section, we explore the numerical values of this timescale for different WDs. However, it is easy to notice from the above equation that the Palatini gravity effect on the cooling time can be negligible in such a simplified model. It is so because the only direct modification is given by $\Phi^{2/7}\approx 1$, while indirect one is hidden in the value of the mass $M$; notice that WDs with limiting masses larger (or smaller) than the Chandrasekhar limit exist only for $\alpha\neq 0$. See the discussion in the following section as well as Figure \ref{Fig: MR}.

\section{Mass--radius relation and cooling age of white dwarfs}\label{Sec4}

Let us firstly solve the hydrostatic balance equations to obtain the mass--radius relation of the WDs. We perform these calculations numerically and present them as the mass--radius curves, which will be helpful for analyzing our cooling model. In our previous paper~\cite{2022PhRvD.105b4028S}, we already obtained these curves for both the Einstein and Jordan frames, and showed that in a WD system, the results do not differ significantly. It is so because in the case of non-relativistic equations, the additional contribution enters via functions of $\Phi$, which does not differ a lot from $1$ for the low-density regimes; that is, $\Phi\approx 1$, in the considered Palatini model. However, at high-density regimes, it has some important consequences.
\begin{figure}[htpb]
	\centering
	\includegraphics[scale=0.55]{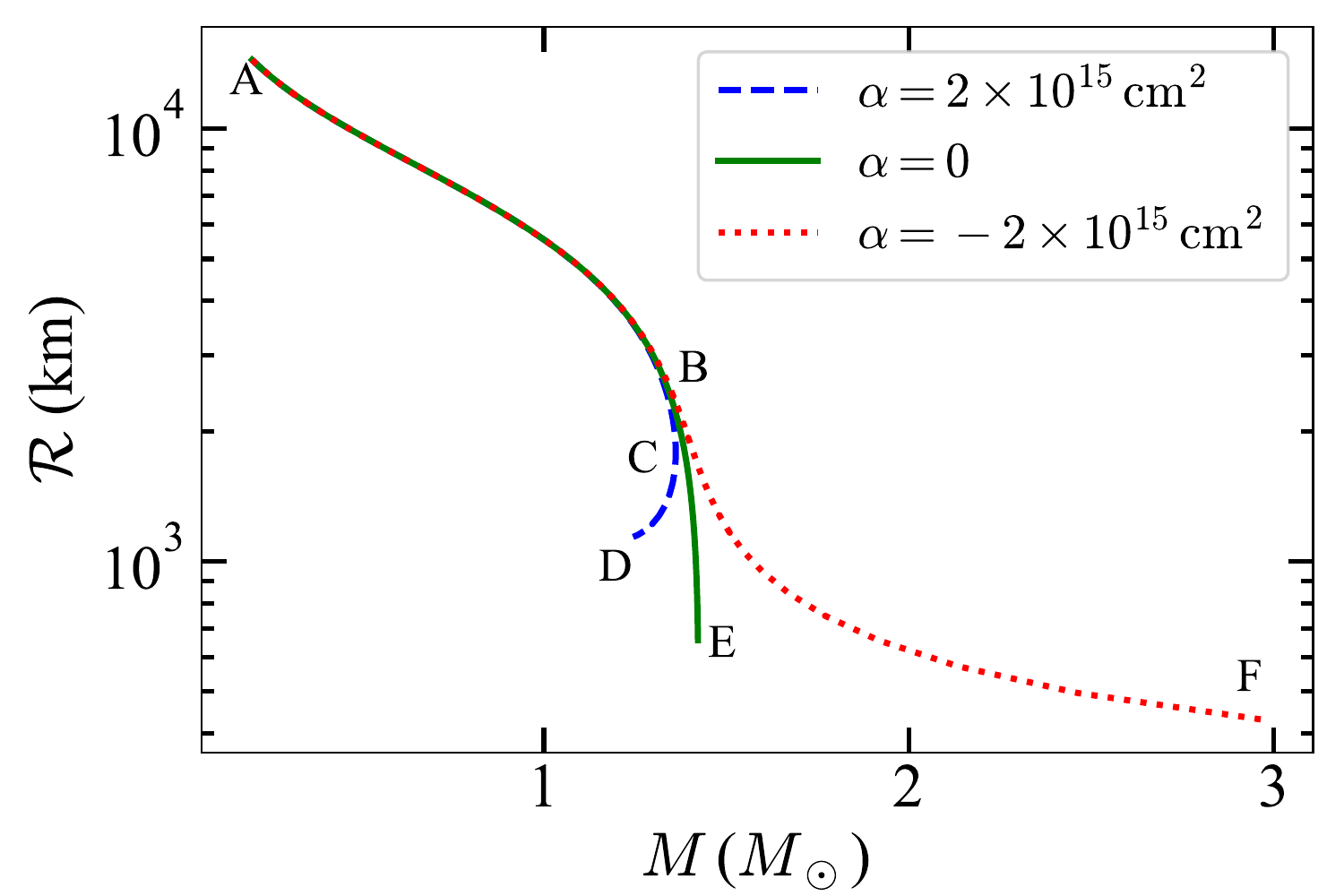}
	\caption{Mass--radius relation of $f(R)$ gravity inspired WDs. The GR case is represented by $\alpha=0$.}
	\label{Fig: MR}
\end{figure}

We solve Equations~\eqref{Eq: dPdr Jordan} and~\eqref{Eq: dmdr Jordan} simultaneously along with the Chandrasekhar EoS with the boundary conditions being $m(r=0)=0$, $\rho(r=0)=\rho_\mathrm{c}$, $m(r=\mathcal{R})=M$, and $\rho(r=\mathcal{R})=0$, where $M$ is the mass of a WD with radius $\mathcal{R}$ and central density $\rho_\mathrm{c}$. Figure~\ref{Fig: MR} depicts the mass--radius curves of WDs for different values of $\alpha$. Note that $\alpha=0$ is the Chandrasekhar result, with mass-limit being approximately $1.44\,M_\odot$. Modified gravity is prominent only in the high-density regime. Hence, the AB branch overlaps with the Chandrasekhar original mass--radius curve for $\alpha=\pm 2\times10^{15}\rm\,cm^2$. Beyond point~B, the curves differ from the original curve. For $\alpha>0$, the curve reaches a maximum mass at point~C and then turns back. It was already shown that the CD branch is unstable under radial perturbation, and hence point~C corresponds to the limiting mass, which turns out to be sub-Chandrasekhar. On the other hand, for $\alpha<0$, the curve turns and reaches the point~F. This BF branch is more stable than the BE branch under radial perturbation, and hence points lying in this branch represent the super-Chandrasekhar WDs. Thus, positive and negative values of the parameter $\alpha$ can respectively explain the existence of sub- and super-Chandrasekhar limiting mass WDs. Let us also note that more significant deviations are expected in the more realistic description, with improved interior's and atmosphere's modelling.

\begin{figure}[htpb]
	\centering
	\includegraphics[scale=0.55]{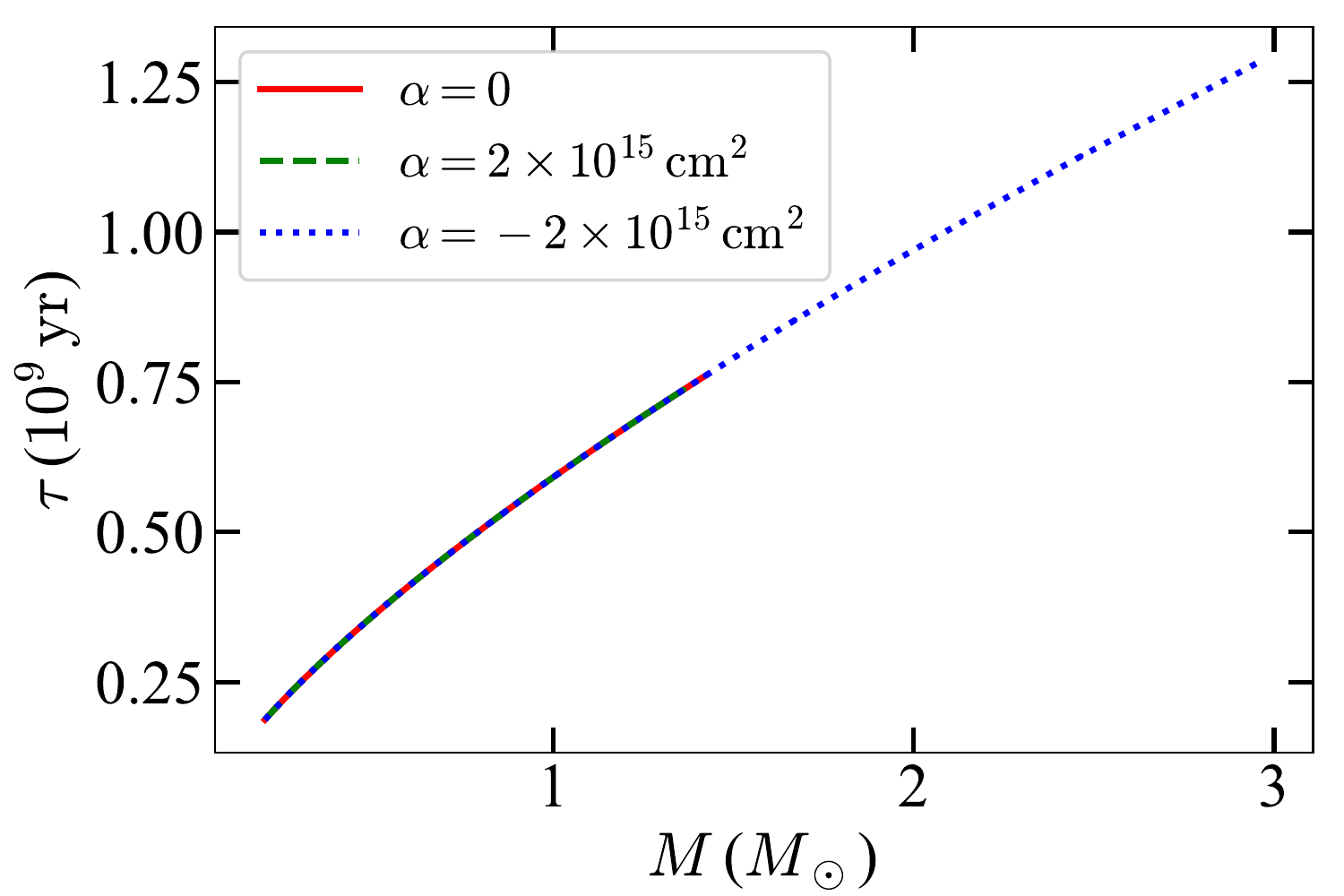}
	\caption{Cooling timescale as a function of the mass of the WDs for $L_*=10^{-3}L_\odot$. Notice that modified gravity allows the white dwarf stars to have higher masses than the Chandrasekhar limit.}
	\label{Fig: age1}
\end{figure}
\begin{figure}[htpb]
	\centering
	\includegraphics[scale=0.55]{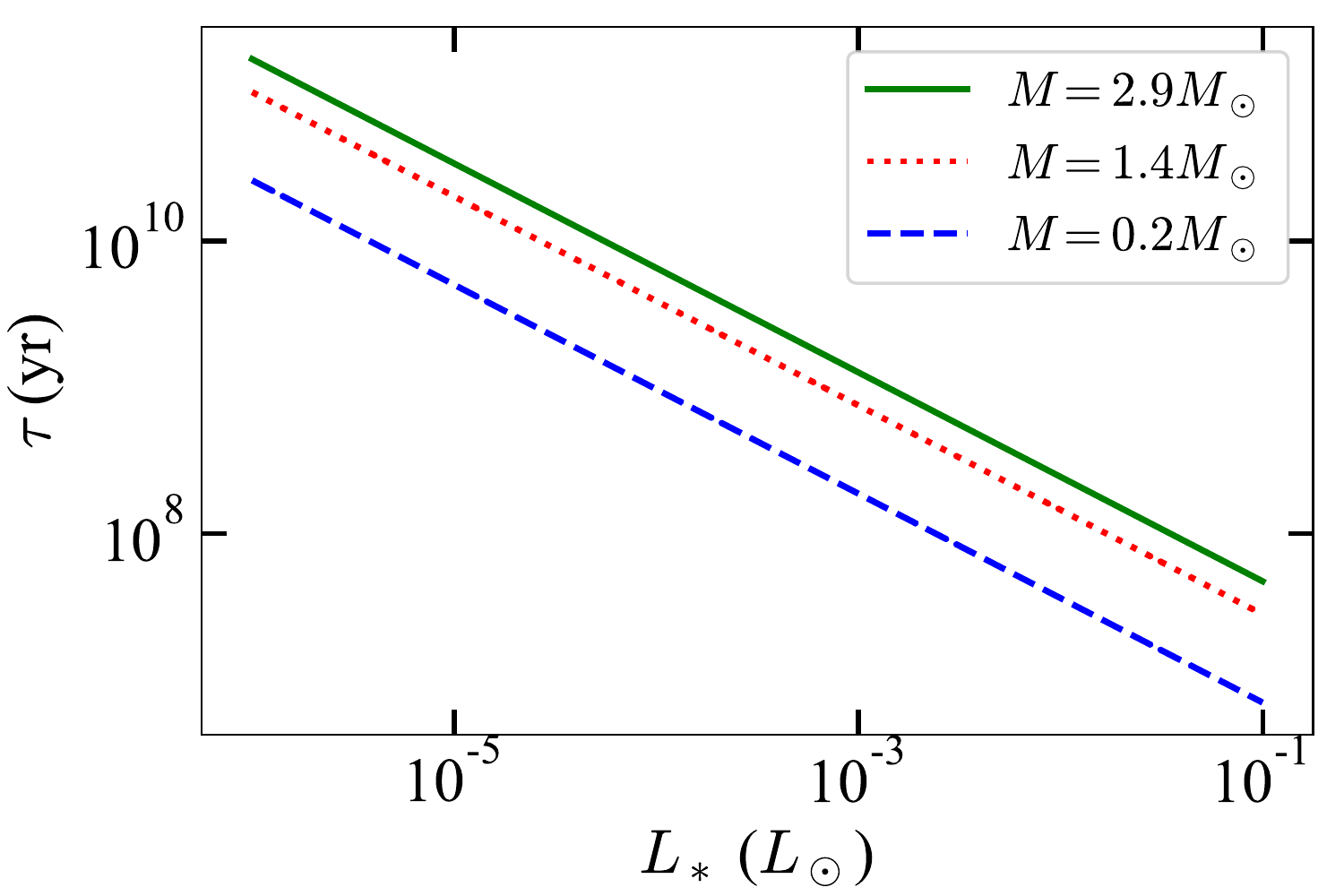}
	\caption{Cooling timescale as a function of the luminosity for WDs with different masses. Note that $2.9\,M_\odot$ WD is possible only when $\alpha<0$. It is worth mentioning that $\alpha$ value is not required explicitly to obtain this plot, because from Figure~\ref{Fig: age1}, it is evident that for a specific luminosity, cooling timescale depends only on the mass, and not the radius. Moreover, the $0.2\,M_\odot$ WD can be obtained for regardless of $\alpha$ being positive, negative, or zero, and hence it is not necessary to precisely specify a value of $\alpha$ in order to determine its cooling timescale.}
	\label{Fig: age2}
\end{figure}

Let us now explore the cooling timescale for sub- and super-Chandrasekhar WDs, and compare them with the conventional ones for the same $\rho_\mathrm{c}$. We know $\Phi = 1 - 16\pi G \rho/c^2$, and hence near the surface, where $\rho\approx0$, we have $\Phi\approx1$. Thus from Equation~\eqref{Eq: age3}, we have $\tau \propto T_*^{-5/2}$, which means if the surface temperatures of the WDs are almost the same, then their cooling timescale remains indistinguishable irrespective of whether be it a sub- or super- or conventional WD, provided their chemical compositions are identical. Figure~\ref{Fig: age1} shows the variation of $\tau$ as a function of $M$ for the luminosity $L_*=10^{-3}L_\odot$. Since $\tau\propto M^{5/7}$, it is evident that as mass increases, the star's age also does it. Thus, it is expected that the cooling timescale is longer for the super-Chandrasekhar WDs and shorter for the sub-Chandrasekhar WDs than conventional ones. However, because $L\propto M$ and $\tau$ depends only on the surface temperature in this simple model, we find that the ages are indistinguishable for the same masses for all the three mass--radius tracks. Moreover, Figure~\ref{Fig: age2} shows the cooling timescale as a function of the luminosity for WDs with different masses. We can see that the higher the mass of the WDs, the longer the cooling timescale of the WD. Because $\tau\propto L^{-5/7}$, this timescale decreases with the increase in the luminosity. This is so because, with the higher luminosity, a WD can quickly release its internal energy and becomes cold. In summary, cooling is faster for heavier WDs with a larger surface luminosity.

\section{Conclusions}\label{Sec5}
The present study has been designed to determine the effect of modifications to the Newtonian hydrostatic equilibrium equation, introduced by the Palatini $f(R)$ gravity, on the cooling model of white dwarf stars. Since it is the first study related to these topics, it was limited to the set of assumptions and simplifications. Firstly, we have assumed that such an object can be modelled as a spherical-symmetric ball of degenerate electrons. It allows us to describe its interior with an analytical equation of state. The second approximation is related to the opacity, which we have assumed to be Kramer's one. Because it also has an analytical form, we could carry our calculations without the need of simulations, or without more advanced numerical methods. Using subsequently the temperature gradient expression, we were able to integrate the hydrostatic balance equation. Applying the ideal gas equation of state, which describes well the behavior of the particles on the surface in our modelling, we have derived the luminosity as a function of mass, temperature, and chemical composition in the presence of modified gravity, which is given by Equation~\eqref{lum1}. As already mentioned, all those considerations are made in the non-relativistic limit of the theory. Moreover, in such a simple model, it is assumed that there is no energy generation in the stars' interior (such as, the latent heat, which is a product of the crystallization process, or further gravitational contraction). Therefore, the white dwarf radiates the whole stored energy away, which means that it is cooling down with time. This fact allows us to write the expression for the cooling timescale as a function of the star's mass and its luminosity as provided by Equation~\eqref{Eq: age3}.

Recalling our previous results on the stability with similar assumptions on the white dwarf model allowed us to immediately notice that we do not expect any significant differences in our cooling for the low-density regime~\cite{2022PhRvD.105b4028S}. We have found there that $\alpha>0$ gives the sub-Chandrasekhar limiting mass white dwarfs and $\alpha<0$ gives the super-Chandrasekhar ones. It seems that among the various theories and models, modified gravity provides a mechanism which is able to justify both mass ranges self-consistently, which can further explain the origins of peculiar under- and and over-luminous type Ia supernovae. Let us now discuss how the previous work helps us to interpret the current findings. 

In this paper, we have found that the cooling timescale is longer for the super-Chandrasekhar white dwarfs than the Newtonian or sub-Chandrasekhar ones. Note that independently of the mass of the dead star, varying the parameter in the given range does not provide any significant differences in the cooling timescale. However, the very massive ones can exist only in modified gravity\footnote{Of course, there are other processes, such as magnetic field~\cite{2013PhRvL.110g1102D,2019MNRAS.490.2692K}, noncommutative geometry~\cite{2021IJMPD..3050034K,2021IJMPD..3050101K}, ungravity eﬀect~\cite{2016PhRvD..93j4046B}, consequence of total lepton number violation~\cite{2015NuPhA.937...17B}, generalized Heisenberg uncertainty principle~\cite{2018JCAP...09..015O}, and many more which can also explain massive white dwarfs but they are failing to explain the sub-Chandrasekhar mass-limit.}. It can be surprising because in the less compact objects, such as the main-sequence stars, brown dwarfs, and even planets~\cite{2020PhRvD.102l4045W,2021PhRvD.103d4037W,Wojnar:2022ttc}, the effects of Palatini gravity manifest in the stellar evolution, even in the cooling processes~\cite{2021PhRvD.103f4032B,2021PhRvD.104j4058W}. This result has however an explanation: the physics we used in our modelling is too poor with respect to the models of non-relativistic stars. In the mentioned works, the authors employed much more realistic equation of state, which included information not only on the electron degeneracy, but also how it evolves with time. This is a theory-dependent process, too~\cite{2021PhRvD.103f4032B,Kozak:2022hdy}. These models took also into account the corrections related to the finite temperature of the gas, mixtures of highly degenerate fluid with the ideal one, as well as ionized elements, and phase transitions in some cases~\cite{auddy2016analytic}. Including the missing physics in order to describe a realistic white dwarf, we also expect to see differences with respect to the Newtonian or GR description. As the most important improvement we would like to analyze is taking into account crystallization processes and effects of gravity on matter properties~\cite{2022arXiv220804023W}. This is, however, out of the scope of this paper. We will present results along these lines in the near future. 

\section*{Acknowledgements}
SK would like to acknowledge support from the South African Research Chairs Initiative of the Department of Science and Technology and the National Research Foundation. AW was supported by the EU through the European Regional Development Fund CoE program TK133 ``The Dark Side of the Universe.'' 

\begin{adjustwidth}{-\extralength}{0cm}

\reftitle{References}


\bibliography{biblio}

%


\end{adjustwidth}
\end{document}